\pdfoutput=1
\documentclass[english]{spie}
\usepackage[T1]{fontenc}
\usepackage[latin9]{inputenc}
\usepackage{refstyle}
\usepackage{units}
\usepackage{textcomp}
\usepackage{amsmath}
\usepackage{graphicx}

\makeatletter

\AtBeginDocument{\providecommand\tabref[1]{\ref{tab:#1}}}
\AtBeginDocument{\providecommand\eqref[1]{\ref{eq:#1}}}
\providecommand{\tabularnewline}{\\}
\RS@ifundefined{subsecref}
  {\newref{subsec}{name = \RSsectxt}}
  {}
\RS@ifundefined{thmref}
  {\def\RSthmtxt{theorem~}\newref{thm}{name = \RSthmtxt}}
  {}
\RS@ifundefined{lemref}
  {\def\RSlemtxt{lemma~}\newref{lem}{name = \RSlemtxt}}
  {}

\usepackage{adjustbox}
\usepackage{siunitx}
\usepackage{array}

\usepackage[labelsep=period,font=footnotesize,margin=0.4in]{caption}

\makeatother

\usepackage{babel}
\begin{document}

\title{Study of higher-order correlation functions and photon statistics
using multiphoton-subtracted states and quadrature measurements\thanks{\enskip{}Citation: Yu. I. Bogdanov, K. G. Katamadze, G. V. Avosopyants,
L. V. Belinsky, N. A. Bogdanova, S. P. Kulik, V. F. Lukichev Study
of higher order correlation functions and photon statistics using
multiphoton-subtracted states and quadrature measurements // Proc.
SPIE 10224, International Conference on Micro- and Nano-Electronics
2016, 102242Q (December 30, 2016)}}

\author{Yu.I.~Bogdanov,\hspace*{-0.18em}\supit{a,b,c} K.G.~Katamadze,\hspace*{-0.18em}\supit{a,b,d}
G.V.~Avosopyants,\hspace*{-0.18em}\supit{a,c,d} L.V.~Belinsky,\hspace*{-0.18em}\supit{a,c}
N.A.~Bogdanova,\hspace*{-0.18em}\supit{a,c} S.P.~Kulik,\hspace*{-0.18em}\supit{d}
V.F.~Lukichev\supit{a}}

\authorinfo{\noindent Further author information:\\
Yu.I.B.: E-mail: bogdanov\_yurii@inbox.ru\\
L.V.B.:\, E-mail: belinsky.leonid@gmail.com}

\maketitle
\noindent \supit{a}Institute of Physics and Technology, Russian Academy
of Sciences, 117218, Moscow, Russia;\\
\supit{b}National Research Nuclear University ``MEPhI'', 115409,
Moscow, Russia;\\
\supit{c}National Research University of Electronic Technology MIET,
124498, Moscow, Russia;\\
\supit{d}M.V. Lomonosov Moscow State University, 119991, Moscow,
Russia
\begin{abstract}
The estimation of high order correlation function values is an important
problem in the field of quantum computation. We show that the problem
can be reduced to preparation and measurement of optical quantum states
resulting after annihilation of a set number of quanta from the original
beam. We apply this approach to explore various photon bunching regimes
in optical states with gamma-compounded Poisson photon number statistics.
We prepare and perform measurement of the thermal quantum state as
well as states produced by subtracting one to ten photons from it.
Maximum likelihood estimation is employed for parameter estimation.
The goal of this research is the development of highly accurate procedures
for generation and quality control of optical quantum states. 
\end{abstract}


\section{Introduction}

Preparation and measurement of optical quantum states are key problems
in applied quantum information technologies. Among the currently used
states of light, the thermal state plays a special role. It serves
as a testbed for various effects based on quantum and classical correlations,
while being easy to prepare. 

The pioneering work of Brown and Twiss \cite{brown_correlation_1956},
which is considered to be the first quantum optics experiment, explored
correlation in thermal light using a beam splitter and a coincidence
circuit with two detectors. Since then thermal states have been used
in many applications including ghost imaging \cite{gatti_ghost_2004,ferri_high-resolution_2005,valencia_two-photon_2005},
quantum illumination \cite{lloyd_enhanced_2008}, and \textquotedblleft thermal
laser\textquotedblright{} \cite{chekhova_intensity_1996}. A recent
demonstration of classical teleportation \cite{guzman-silva_demonstration_2016}
also used thermal states. In this paper we consider the properties
of photon number statistics and autocorrelation functions in photon-subtracted
thermal states. 

Photon addition and subtraction is of the great interest in quantum
optics, because it provides a tool for direct tests of basic commutation
relations \cite{parigi_probing_2007} and enables the preparation
of Schrodinger cat and other exotic quantum states \cite{wenger_non-gaussian_2004}.
It can also be used for probabilistic linear no-noise amplification
\cite{xiang_heralded_2010}. One- and two-photon subtracted thermal
states were demonstrated for the first time in \cite{zavatta_subtracting_2008}.
The measurement of photon statistics with photon number resolving
detectors was demonstrated in \cite{zhai_photon-number-resolved_2013}.
In the present work we provide a comprehensive description of multiphoton
subtracted thermal states, based on a general approach, suitable for
any photon number distribution. We demonstrate the technique of high-fidelity
preparation and reconstruction of up to 10-photon subtracted thermal
states, using detectors not capable of resolving the photon numbers.
The technique discussed in the paper can also be used in some metrological
applications \cite{parazzoli_enhanced_2016,rafsanjani_interferometry_2016}. 

\section{Probability generating functions and autocorrelation}

The approach is based on exploiting the properties of generating functions.
Probability generating functions $G\left(z\right)$ contain all information
about the random distribution of photons. In particular, the probability
of detecting $k$ photons is the derivative of order $k$ evaluated
at zero ${G^{\left(k\right)}}\left(0\right)$. The factorial moment
of order $m$ $E\left[{k\left({k-1}\right)...\left({k-m+1}\right)}\right]$
is equal to the $m$-th order derivative evaluated at $z=1:$ ${G^{\left(m\right)}}\left(1\right).$ 

\begin{equation}
P\left(k\right)=\frac{{{G^{\left(k\right)}}\left(0\right)}}{{k!}};
\end{equation}
\begin{equation}
E\left[{k\left({k-1}\right)...\left({k-m+1}\right)}\right]={G^{\left(m\right)}}\left(1\right),
\end{equation}
where $E$ stands for the expected value. 

Autocorrelation function of order $m,$ which can be measured in an
experiment with $m$ photon detectors, is determined by factorial
moments, and thus can be expressed in terms of the generating function
derivative at $z=1.$

\begin{equation}
{g^{\left(m\right)}}=\frac{{{G^{\left(m\right)}}\left(1\right)}}{{\mu^{m}}},\;m=1,2,...
\end{equation}
Here $\mu=G^{\left(1\right)}\left(1\right)$ is the mean number of
photons in the initial state. This implies the equality $g^{\left(1\right)}=1.$

We use a beam splitter with low reflection probability $p$ to separate
individual photons from the beam. Using the events at the subtracted
photon detector, we select the data pertaining only to the states
from which we have annihilated a photon; everything else is discarded.
Let the initial distribution of the number of photons have the probability
mass function (pmf) $P\left(k\right).$ Taking into account the probability
of exactly one photon being split off at the beam splitter $kp{\left(1-p\right)}^{k-1},$
we have the probability of obtaining the state with $k-1$ photons
equal to $N\cdot P\left(k\right)kp{\left(1-p\right)}^{k-1},$ where
$N$ is the normalisation constant. Thus the pmf of the initial state\textquoteright s
photon number distribution is modified by a factor of $kp{\left(1-p\right)}^{k-1}.$
As $p\rightarrow0$ the factor goes to $k,$ since $p{\left(1-p\right)}^{k-1}\rightarrow1.$
We can express photon subtraction using creation and annihilation
operators: 
\begin{equation}
\rho^{out}=Na\rho^{in}a^{\dagger}.
\end{equation}

Based on this analysis we can show that, as the reflection probability
goes to zero $p\rightarrow0,$ the generating function $G_{1}\left(z\right)$
of the photon number distribution for the photon-subtracted state
is determined by the derivative of the generating function of the
initial state\textquoteright s distribution: 

\begin{equation}
{G_{1}}\left(z\right)=\frac{{{G^{\left(1\right)}}\left(z\right)}}{{{G^{\left(1\right)}}\left(1\right)}}=\frac{{{G^{\left(1\right)}}\left(z\right)}}{\mu}.\label{eq:PGF_derivative_p0}
\end{equation}

In cases where $p$ is not small enough to ignore the above expression
becomes:

\begin{equation}
{G_{1}}\left(z\right)=\frac{{{G^{\left(1\right)}}\left({z\left({1-p}\right)}\right)}}{{{G^{\left(1\right)}}\left({1-p}\right)}}.
\end{equation}

Repeated application of (\ref{eq:PGF_derivative_p0}) allows us to
derive an expression for the probability generating function $G_{m}\left(z\right)$
of a state from which $m$ photons have been subtracted: 

\begin{equation}
{G_{m}}\left(z\right)=\frac{{{G^{\left(m\right)}}\left(z\right)}}{{\mu{\mu_{1}}\cdot\cdot\cdot{\mu_{m-1}}}},\;m=1,2,...
\end{equation}

Here $\mu_{i}$ is the mean number of photons for the state with $i$
subtracted quanta. Iterated measurement of these quantities allows
us to compute autocorrelation functions of arbitrary order:

\begin{equation}
{g^{\left(m\right)}}=\frac{{{\mu_{1}}{\mu_{2}}\cdot\cdot\cdot{\mu_{m-1}}}}{{\mu^{m-1}}},\;m=2,3,...\label{eq:autocorr_order_m}
\end{equation}

In particular, the above reply implies that the second order correlation
function is equal to the ratio of mean photon number in the state
with one photon subtracted $\mu_{1}$ to the mean photon number in
the original state $\mu:$ 

\begin{equation}
{g^{\left(2\right)}}=\frac{{\mu_{1}}}{\mu}.
\end{equation}

In general, the following recurrence allows us to calculate the autocorrelation
function of order $m+1$ from the function of order $m:$ 

\begin{equation}
{g^{\left({m+1}\right)}}={g^{\left(m\right)}}\frac{{\mu_{m}}}{\mu},\;m=1,2,...
\end{equation}

It is important to note that the autocorrelation characteristics of
the conditional states obtained by subtracting quanta can be expressed
in terms of correlations of the original state. The correlation function
$g_{m}^{\left(n\right)}$ of order $n$ for the distribution of the
$m$-subtracted state can be expressed through the correlation functions
of the original distribution: 

\begin{equation}
g_{m}^{\left(n\right)}=\frac{{G_{m}^{\left(n\right)}\left(1\right)}}{{\mu_{m}^{n}}}=\frac{{{g^{\left({m+n}\right)}}{{\left({g^{\left(m\right)}}\right)}^{n-1}}}}{{{\left({g^{\left({m+1}\right)}}\right)}^{n}}},\;n=1,2,...;\,m=1,2,...
\end{equation}

\section{Compound Poisson distribution }

Poisson distribution with parameter $\lambda$ has the probability
generating function: 

\begin{equation}
{G_{0}}\left({z\left|\lambda\right.}\right)=\exp\left({-\lambda\left({1-z}\right)}\right).\label{eq:Poisson_PGF}
\end{equation}

Let the parameter $\lambda$ be a random variable with Gamma probability
distribution with the probability density function (pdf) 

\begin{equation}
P\left(\lambda\right)=\frac{{{b^{a}}{\lambda^{a-1}}{e^{-b\lambda}}}}{{\Gamma\left(a\right)}},
\end{equation}
where parameters $a>0,\,b>0.$ And $\Gamma\left(a\right)$ is the
gamma function. 

The resulting compound distribution has the following probability
generating function:

\begin{equation}
G\left({z\left|{a,b}\right.}\right)=\int\limits _{0}^{\infty}{{G_{0}}\left({z\left|\lambda\right.}\right)P\left(\lambda\right)}d\lambda=\frac{1}{{{\left({1+\frac{{\left({1-z}\right)}}{b}}\right)}^{a}}}.
\end{equation}

This compound distribution for positive values of $a$ is known as
the negative binomial distribution. It has the mean of $\mu=\nicefrac{a}{b}$
and can be reparameterised as: 

\begin{equation}
G\left({z\left|{\mu,a}\right.}\right)=\frac{1}{{{\left({1+\frac{{\mu\left({1-z}\right)}}{a}}\right)}^{a}}}.\label{eq:compoud_PGF_mu_a}
\end{equation}

Thus the gamma-compouded Poisson distribution has two parameters:
$\mu$ is the mean number of photons and $a$ is the photon clusterisation
factor. This parameter can also be interpreted as the degree of coherency:
as $a$ rises, the photon distribution converges to a Poisson distribution
with the same mean, which gives photon number distribution of the
coherent state. It is easy to see that (\ref{eq:compoud_PGF_mu_a})
is equal to the generating function of the Poisson distribution in
the $a\rightarrow\infty$ limit. The case of $a=1$ corresponds to
the thermal state. 

This distribution describes a multimode thermal state, where $a$
is the number of modes \cite{mandel1995optical}. It can be shown
that the same distribution also applies to the single-mode multiphoton-subtracted
thermal state \cite{BogdanovAvtometr2016}. 

We note that taking optical losses into account corresponds to a simple
scaling of the mean photon number: $\mu\rightarrow\mu t,$ where $t$
is the transmission coefficient: $t=1-\gamma,$ with $\gamma$ equal
to the absorbed proportion of energy. 

Another interesting fact is that the distribution generated by (\ref{eq:compoud_PGF_mu_a})
is well defined not only for the positive values of $a,$ but also
for negative integers: $a=-n,$ where $n=1,2,\ldots.$ Such distributions
can be used to model states with $n$ photons as long as the condition
$0<\mu\le n$ is satisfied. The case $\mu=n$ corresponds to the case,
where the prepared state contains precisely $n$ photons. In case
of $\mu<n,$ we can interpret the distribution as describing the situation
after each of $n$ photos was subject to annihilation with probability
of survival $\theta=\frac{\mu}{n}.$ The resulting binomial distribution
with probability mass function $f\left(k;n,\theta\right)$ then gives
the probability of $k$ photons surviving, out of initial set of $n.$ 

Using the results from section 2, it is easy to show that autocorrelation
function of order $m$ for the gamma-compounded Poisson is given by: 

\begin{equation}
{g^{\left(m\right)}}=\frac{{{\left(a\right)}_{m}}}{{a^{m}}}=\frac{{a\left({a+1}\right)...\left({a+m-1}\right)}}{{a^{m}}}.
\end{equation}
Here $\left(a\right)_{k}$ is the rising factorial: 
\begin{equation}
{\left(a\right)_{k}}=a\left({a+1}\right)...\left({a+k-1}\right).
\end{equation}

In particular: ${g^{\left(2\right)}}=\frac{{a+1}}{a},$ ${g^{\left(3\right)}}=\frac{{\left({a+1}\right)\left({a+2}\right)}}{{a^{2}}},$
${g^{\left(4\right)}}=\frac{{\left({a+1}\right)\left({a+2}\right)\left({a+3}\right)}}{{a^{3}}}.$

For example a thermal state has $a=1$ and $g^{\left(2\right)}=2;$
one-photon Fock state has $a=-1$ and $g^{\left(2\right)}=0$ and
a coherent state has $g^{\left(2\right)}=1$ with $a\rightarrow\infty.$

Using (\ref{eq:PGF_derivative_p0}) it is easy to show that the conditional
state obtained after photon subtraction has the following important
property: the photon number distribution keeps the compound Poisson
type, the parameter $b$ is unchanged and the parameter $a$ is simply
incremented by one: 
\begin{equation}
a^{\left(1\right)}=a+1.
\end{equation}

From the above we derive the expression for the mean number of photons
after subtraction: 
\begin{equation}
{\mu^{\left(1\right)}}=\frac{{\mu\left({a+1}\right)}}{a}.
\end{equation}
Here $a^{\left(1\right)}$ and $\mu^{\left(1\right)}$ are the parameters
of the photon-subtracted state\textquoteright s distribution.

After subtracting $m$ photons we obtain: 
\begin{equation}
{a^{\left(k\right)}}=a+k;
\end{equation}

\begin{equation}
{\mu^{\left(k\right)}}=\frac{{\mu\left({a+k}\right)}}{a}.
\end{equation}

In case of a thermal source $a=1,$ so 
\begin{equation}
{a^{\left(k\right)}}=1+k,\quad{\mu^{\left(k\right)}}=\mu\left({1+k}\right).
\end{equation}

In particular $\mu^{\left(1\right)}=2\mu$ and after subtracting one
photon we have raised the expected number of photons in the state
by a factor of two. This increase does not represent a paradox, since
the increased photon count is expected in the conditional state\textquoteright s
distribution. Subtracting a photon lowers the photon count by one,
but at the same time the subtraction event is more likely for the
states with higher photon counts in the original state. In case of
thermal states, the second effect is powerful enough to double the
expected number of photons. 

\section{Quadrature measurements of states }

In case of the mixture with the compound-Poisson probability distribution
$P\left({k\left|{\mu,a}\right.}\right)$ over Fock components we have
the following density matrix in the in-phase quadrature basis: 
\begin{equation}
\rho\left({x,x'}\right)=\sum\limits _{k=0}^{\infty}{P\left({k\left|{\mu,a}\right.}\right)}{\varphi_{k}}\left(x\right)\varphi_{k}^{*}\left({x'}\right)
\end{equation}
Here $\varphi_{k}\left(x\right),\;k=0,1,2,\ldots$ are the Hermite
functions, which form the eigenbasis of the harmonic oscillator. These
functions can be written in closed form as
\begin{equation}
{\varphi_{k}}\left(x\right)=\frac{1}{{{\left({{2^{k}}k!\sqrt{\pi}}\right)}^{1/2}}}{H_{k}}\left(x\right)\exp\left({-\frac{{x^{2}}}{2}}\right),\;k=0,1,2,...
\end{equation}
where $H_{k}$ are the Hermite polynomials.

The corresponding probability distribution over the quadratures is
then given by 
\begin{equation}
P\left({x\left|{\mu,a}\right.}\right)=\sum\limits _{k=0}^{\infty}{P\left({k\left|{\mu,a}\right.}\right)}{\left|{{\varphi_{k}}\left(x\right)}\right|^{2}}.\label{eq:Quadrature_pdf}
\end{equation}

This is an even function, so all the odd moments are equal to zero. 

The variance of the quadrature distribution (\ref{eq:Quadrature_pdf})
does not depend on parameter $a$ and is completely determined by
the mean photon count: 
\begin{equation}
{\sigma^{2}}=\mu+\frac{1}{2}.
\end{equation}

It is also possible to obtain closed form expressions for higher order
moments. In particular the skewness $\beta_{1}$ is equal to zero
and the excess kurtosis $\beta_{2}$ is given by 
\begin{equation}
{\beta_{2}}=-6{\left({\frac{\mu}{{2\mu+1}}}\right)^{2}}\frac{{\left({a-1}\right)}}{a}.
\end{equation}

The above quantities for a random variable $x$ are defined as 
\begin{equation}
{\beta_{1}}=\frac{{E\left[{{\left({x-E\left(x\right)}\right)}^{3}}\right]}}{{\sigma^{3}}},
\end{equation}
\begin{equation}
{\beta_{2}}=\frac{{E\left[{{\left({x-E\left(x\right)}\right)}^{4}}\right]}}{{\sigma^{4}}}-3.
\end{equation}
Here $E$ stands for the expected value and $\sigma$ is the standard
deviation.

It is possible to estimate the mean photon number $\mu$ and the clusterisation
factor $a$ of the mixture from the variance and excess kurtosis of
the quadrature data: 

\begin{equation}
\mu={\sigma^{2}}-\frac{1}{2},
\end{equation}
\begin{equation}
a=\frac{{6{\mu^{2}}}}{{6{\mu^{2}}+{\beta_{2}}{{\left({2\mu+1}\right)}^{2}}}}.
\end{equation}

These formulae give the estimates of the parameters $\mu$ and $a,$
according to the method of moments. We use these estimates as starting
points for a more powerful maximum likelihood estimation procedure. 

We have prepared and measured eleven states: the thermal state and
the photon-subtracted states with the number of subtracted photons
$m$ varying from 1 to 10. The results are presented in \tabref{Results-of-statistical}.
Here we denote the number of subtracted photons as $m$ with $m=0$
referring to the original thermal state. The error bounds are expressed
using standard deviation of parameters. Large bounds for the cases
with nine and ten subtracted photons are a consequence of a small
sample size. Fidelity was calculated between the reconstructed state
and the ideal theoretical state. $\chi^{2}$ significance test was
used to check for consistency between the obtained tomographic models
and the data. In our case all of the significance levels are higher
than one percent and the models are consistent with the data.

\begin{table}
\noindent \centering{}\caption{Results of statistical reconstruction from experimental data\label{tab:Results-of-statistical}}
\begin{tabular}{crrrrc}
\multicolumn{1}{c}{State} &
\multicolumn{1}{c}{$\mu\pm\sigma_{\mu}$} &
\multicolumn{1}{c}{$a\pm\sigma_{a}$} &
\multicolumn{1}{c}{Sample size} &
Fidelity &
$\chi^{2}$ significance level,\tabularnewline
\hline 
\hline 
Thermal $m=0$  &
3.034\textpm 0.022 &
0.999\textpm 0.016 &
50000 &
99.999 &
0.370\tabularnewline
\hline 
$m=1$ &
5.983\textpm 0.051 &
1.605\textpm 0.036 &
25000 &
99.659 &
0.112\tabularnewline
\hline 
$m=2$ &
9.063\textpm 0.093 &
2.515\textpm 0.088 &
12500 &
99.777 &
0.550\tabularnewline
\hline 
$m=3$ &
12.261\textpm 0.150 &
3.149\textpm 0.147 &
7500 &
99.572 &
0.014\tabularnewline
\hline 
$m=4$ &
15.538\textpm 0.223 &
4.331\textpm 0.281 &
4500 &
99.799 &
0.565\tabularnewline
\hline 
$m=5$ &
17.957\textpm 0.244 &
5.198\textpm 0.350 &
4500 &
99.831 &
0.111\tabularnewline
\hline 
$m=6$ &
21.050\textpm 0.362 &
6.378\textpm 0.600 &
2500 &
99.928 &
0.214\tabularnewline
\hline 
$m=7$ &
24.732\textpm 0.410 &
7.045\textpm 0.665 &
2500 &
99.832 &
0.021\tabularnewline
\hline 
$m=8$ &
27.795\textpm 0.433 &
8.847\textpm 0.882 &
2500 &
99.945 &
0.207\tabularnewline
\hline 
$m=9$ &
30.536\textpm 0.999 &
11.259\textpm 2.680 &
500 &
99.896 &
0.021\tabularnewline
\hline 
$m=10$ &
33.114\textpm 1.267 &
11.337\textpm 3.156 &
358 &
99.980 &
0.123\tabularnewline
\hline 
\end{tabular}
\end{table}

According to \eqref{autocorr_order_m}, the product of all the mean
photon numbers gives the autocorrelation function of the eleventh
order: ${g^{\left({11}\right)}}=\frac{{\prod\limits _{m=1}^{10}{\mu_{m}}}}{{\mu^{10}}}.$
Using the data from \tabref{Results-of-statistical} we obtain $\ln{g^{\left({11}\right)}}=17.53\pm0.10$.
The theoretical value is equal to $\ln11!=17.50.$ 

\section{The hierarchy of compound Poisson distributions}

In section 3 of the article we let the parameter $\lambda$ of the
Poisson distribution be a random variable with a gamma distribution,
which gives the gamma-compound Poisson distribution. However, this
model can be viewed as the first order approximation, with the original
Poisson corresponding to the zeroth order. If we let the mean photon
count $\mu$ of the compound Poisson be a random variable, we can
continue the process of obtaining successively more sophisticated
models. The physical justification for this operation is as follows.
The stabilisation of the mean photon count inside the exposition time
of $\tau$ is not absolute. In general the different frames are formed
in different conditions and the mean time may differ from one to the
other. The second level model allows us to take this variation into
account. Various groups of frames may also be inhomogeneous, which
would lead to a third level model and so on. 

Following \cite{Bogdanov2003} we derive the expressions for the probability
generating functions of the distributions in this hierarchy. We start
from the first level gamma compound Poisson probability generating
function in the following form: 
\begin{equation}
{G_{1}}\left(z\right)=\frac{1}{{{\left({1+\frac{{\left({1-z}\right)}}{{b_{1}}}}\right)}^{{a_{1}}}}}={\left({1+\frac{{\left({1-z}\right)}}{{b_{1}}}}\right)^{-\mu{b_{1}}}}=\exp\left[{-\mu{b_{1}}\ln\left({1+\frac{{\left({1-z}\right)}}{{b_{1}}}}\right)}\right].\label{eq:Two-level_PGF}
\end{equation}

Now let the mean photon count $\mu$ be a random variable described
by a gamma probability distribution with parameters $a_{2}$ and $b_{2},$
while the $b_{1}$ is constant. In this case we obtain the second
level model. Note that we can transform a zeroth level model (\ref{eq:Poisson_PGF})
into a first level one by means of a simple formal substitution: 
\begin{equation}
\left({1-z}\right)\to{b_{1}}\ln\left({1+\frac{{\left({1-z}\right)}}{{b_{1}}}}\right).
\end{equation}

Averaging over the mean $\mu$ in (\ref{eq:Two-level_PGF}) we obtain
the generating function of the second level compound Poisson model
with parameters $a_{2}$ and $b_{2}$:
\begin{equation}
{G_{2}}\left({z\left|{\mu,{b_{1}},{b_{2}}}\right.}\right)=\frac{1}{{{\left({1+\frac{{b_{1}}}{{b_{2}}}\ln\left({1+\frac{{\left({1-z}\right)}}{{b_{1}}}}\right)}\right)}^{\mu{b_{2}}}}}.
\end{equation}

Here $\mu$ is the general expectation, $b_{1}$ and $b_{2}$ are
parameters of the first and second levels of hierarchy, respectively.
We can also define the clusterisation parameters for the first and
second levels as $a_{1}=\mu b_{1},$ $a_{2}=\mu b_{2}.$ We note that
as $a_{2}\rightarrow\infty,$ $b_{2}\rightarrow\infty,$ while $\nicefrac{a_{2}}{b_{2}}=\mu=const,$
the second level model converges to the first order one. 

Iteratively repeating the above procedure we obtain the following
recurrence relations between probability generating functions of various
levels of the hierarchy:
\begin{gather}
{G_{r}}\left({z\left|{\mu,{b_{1}},...,{b_{r}}}\right.}\right)=\exp\left[{-\mu{L_{r}}}\right],\nonumber \\
{L_{0}}=\left({1-z}\right)\\
{L_{r+1}}={b_{r+1}}\ln\left({1+\frac{{L_{r}}}{{b_{r+1}}}}\right),\;r=0,1,\ldots\nonumber 
\end{gather}

For each of the levels along with parameters $b_{r}$ we can also
define the clusterisation parameters $a_{r}=\mu b_{r},\;r=1,2,\ldots$ 

The experiments demonstrate a relatively small inconsistency of the
data with the first level model and the necessity of employing the
second level model. 

\begin{figure}
\noindent \begin{centering}
\includegraphics[width=0.75\columnwidth]{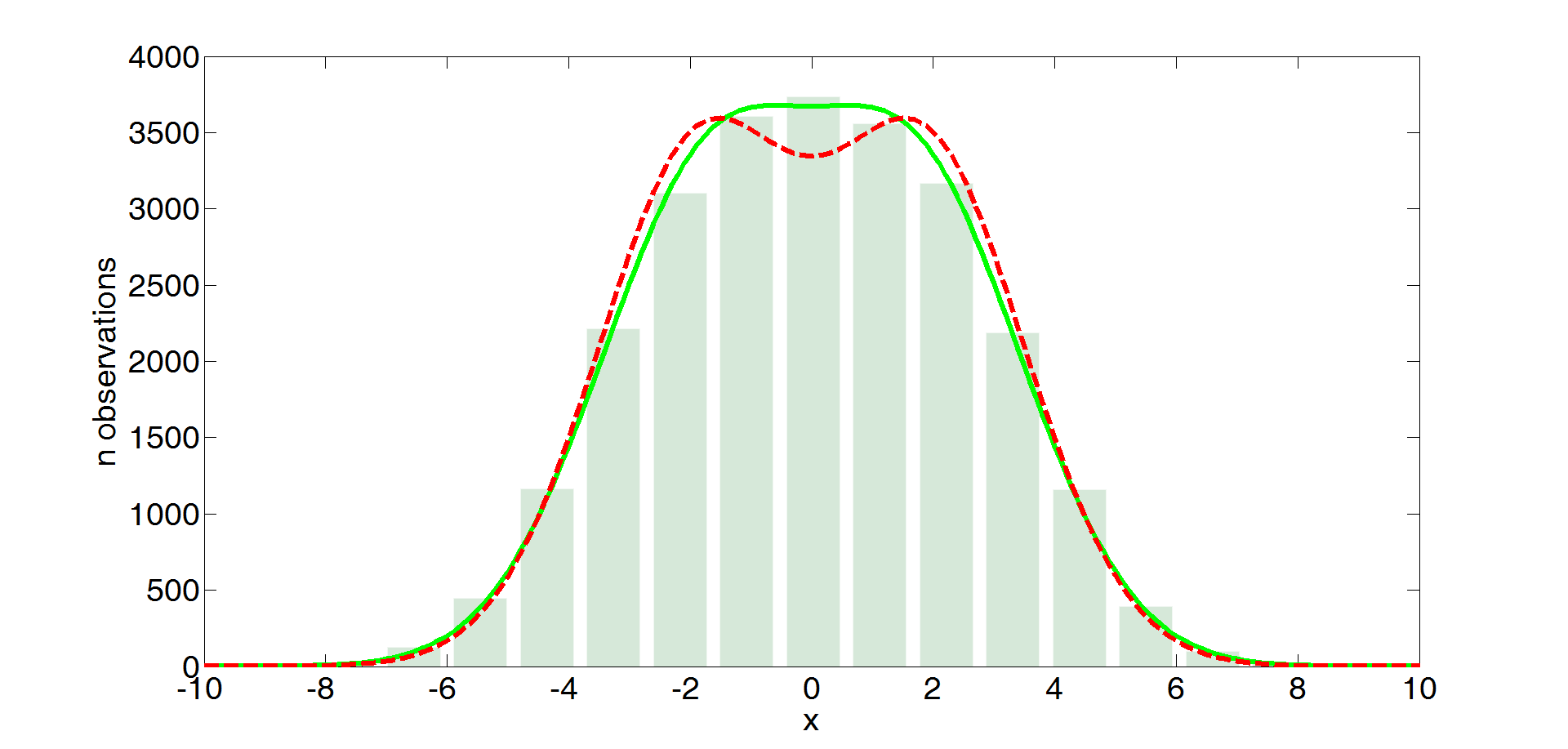}
\par\end{centering}
\caption{Comparison between models for the state with one subtracted photon.
Dashed red line is the first level model with $a^{ideal}=2$ and the
solid green line corresponds to a second level model with $a_{1}=2,$
$a_{2}=8.46.$\label{fig:Comparison-between-models}}

\end{figure}

Figure \ref{fig:Comparison-between-models} depicts the comparison
between models for the state with one subtracted photon.

Here the experimental sample is used to form the histogram, the dashed
red line is the first level model with $a^{ideal}=2$ and the solid
green line corresponds to a second level model with $a_{1}=2,$ $a_{2}=8.46.$
$\mu$ is equal to $5.98$ in both cases. The second level model clearly
fits the data better than the ideal theoretic model with no corrections
(in particular, the corrected model has no minimum at $0$). Note
that the second level corrections tend to zero as the parameter $a_{2}\rightarrow\infty.$

\section{Conclusion}

Photon statistics of a family of photon-subtracted thermal states
have been described using the probability generating functions of
the photon number distribution. The correlation functions of various
orders were expressed in terms of quantities measurable using only
detectors incapable of resolving the number of photons. Up to ten-photon
subtracted states have been experimentally realised and measured with
$>99\%$ fidelity. Our results showcase the flexibility of this model
in analysis of quantum optical experimental data.
\begin{acknowledgments}
The work was supported by the Russian Science Foundation, grant no
14-12-01338.
\end{acknowledgments}
\bibliographystyle{2E__My_Documents_work_docs_articles_SPIE_2016_negBinom_spiebib}
\bibliography{1E__My_Documents_work_docs_articles_SPIE_2016_negBinom_SPIE2016_bibtex_t1}

\begin{thebibliography}{10}

\bibitem{brown_correlation_1956}
Brown, R.~H. and Twiss, R.~Q., ``Correlation between {Photons} in two
  {Coherent} {Beams} of {Light},'' {\em Nature}~{\bf 177},  27--29 (Jan. 1956).

\bibitem{gatti_ghost_2004}
Gatti, A., Brambilla, E., Bache, M., and Lugiato, L.~A., ``Ghost {Imaging} with
  {Thermal} {Light}: {Comparing} {Entanglement} and {ClassicalCorrelation},''
  {\em Physical Review Letters}~{\bf 93},  093602 (Aug. 2004).

\bibitem{ferri_high-resolution_2005}
Ferri, F., Magatti, D., Gatti, A., Bache, M., Brambilla, E., and Lugiato,
  L.~A., ``High-{Resolution} {Ghost} {Image} and {Ghost} {Diffraction}
  {Experiments} with {Thermal} {Light},'' {\em Physical Review Letters}~{\bf
  94},  183602 (May 2005).

\bibitem{valencia_two-photon_2005}
Valencia, A., Scarcelli, G., D'Angelo, M., and Shih, Y., ``Two-{Photon}
  {Imaging} with {Thermal} {Light},'' {\em Physical Review Letters}~{\bf 94},
  063601 (Feb. 2005).

\bibitem{lloyd_enhanced_2008}
Lloyd, S., ``Enhanced {Sensitivity} of {Photodetection} via {Quantum}
  {Illumination},'' {\em Science}~{\bf 321},  1463--1465 (Sept. 2008).

\bibitem{chekhova_intensity_1996}
Chekhova, M.~V., Kulik, S.~P., Penin, A.~N., and Prudkovskii, P.~A.,
  ``Intensity interference in {Bragg} scattering by acoustic waves with thermal
  statistics,'' {\em Physical Review A}~{\bf 54},  R4645--R4648 (Dec. 1996).

\bibitem{guzman-silva_demonstration_2016}
Guzman-Silva, D., Brüning, R., Zimmermann, F., Vetter, C., Gräfe, M., Heinrich,
  M., Nolte, S., Duparré, M., Aiello, A., Ornigotti, M., and Szameit, A.,
  ``Demonstration of local teleportation using classical entanglement,'' {\em
  Laser \& Photonics Reviews}~{\bf 10},  317--321 (Mar. 2016).

\bibitem{parigi_probing_2007}
Parigi, V., Zavatta, A., Kim, M., and Bellini, M., ``Probing {Quantum}
  {Commutation} {Rules} by {Addition} and {Subtraction} of {Single} {Photons}
  to/from a {Light} {Field},'' {\em Science}~{\bf 317},  1890--1893 (Sept.
  2007).

\bibitem{wenger_non-gaussian_2004}
Wenger, J., Tualle-Brouri, R., and Grangier, P., ``Non-{Gaussian} {Statistics}
  from {Individual} {Pulses} of {Squeezed} {Light},'' {\em Physical Review
  Letters}~{\bf 92},  153601 (Apr. 2004).

\bibitem{xiang_heralded_2010}
Xiang, G.~Y., Ralph, T.~C., Lund, A.~P., Walk, N., and Pryde, G.~J., ``Heralded
  noiseless linear amplification and distillation of entanglement,'' {\em
  Nature Photonics}~{\bf 4},  316--319 (May 2010).

\bibitem{zavatta_subtracting_2008}
Zavatta, A., Parigi, V., Kim, M.~S., and Bellini, M., ``Subtracting photons
  from arbitrary light fields: experimental test of coherent state invariance
  by single-photon annihilation,'' {\em New Journal of Physics}~{\bf 10}(12),
  123006 (2008).

\bibitem{zhai_photon-number-resolved_2013}
Zhai, Y., Becerra, F.~E., Glebov, B.~L., Wen, J., Lita, A.~E., Calkins, B.,
  Gerrits, T., Fan, J., Nam, S.~W., and Migdall, A., ``Photon-number-resolved
  detection of photon-subtracted thermal light,'' {\em Optics Letters}~{\bf
  38},  2171 (July 2013).

\bibitem{parazzoli_enhanced_2016}
Parazzoli, C.~G. and Capron, B.~A., ``Enhanced {Thermal} {Images} of {Faint}
  {Objects} via {Photon} {Addition} / {Subtraction},''  FTu3C.4, OSA (2016).

\bibitem{rafsanjani_interferometry_2016}
Rafsanjani, S. M.~H., Mirhosseini, M., Magana-Loaiza, O.~S., Gard, B.~T.,
  Birrittella, R., Koltenbah, B.~E., Parazzoli, C.~G., Capron, B.~A., Gerry,
  C.~C., Dowling, J.~P., and Boyd, R.~W., ``Interferometry with
  {Photon}-{Subtracted} {Thermal} {Light},'' {\em arXiv:1605.05424 [physics,
  physics:quant-ph]}  (May 2016).

\bibitem{mandel1995optical}
Mandel, L. and Wolf, E.,  [{\em Optical Coherence and Quantum
  Optics}{\nolinebreak\hspace{0.1em}]}, Cambridge University Press (1995).

\bibitem{BogdanovAvtometr2016}
Bogdanov, {\relax Yu}.~I., Bogdanova, N.~A., Katamadze, K.~G., Avosopyants,
  G.~V., and Lukichev, V.~F., ``Study of photon statistics with use of compound
  {Poisson} distribution and quadrature measurements,'' {\em Avtometriya}~{\bf
  52}(5),  71--83 (2016).
\newblock [\textit{Optoelectronics, Instrumentation and Data Processingng},
  \textbf{52}(5), 71--83 (in press)].

\bibitem{Bogdanov2003}
Bogdanov, {\relax Yu}.~I., Bogdanova, N.~A., and Dshkhunyan, V.~L.,
  ``Statistical yield modeling for {IC} manufacture: Hierarchical fault
  distributions,'' {\em Russian Microelectronics}~{\bf 32}(1),  51--62 (2003).

\end{thebibliography}

\end{document}